\documentclass{ws-procs975x65}

\usepackage{ifpdf}

\begin{document}

\title{\uppercase{Cosmological constraints on ghost-free bigravity:\\ Background dynamics and late-time acceleration}}

\author{YASHAR AKRAMI$^*$, TOMI S. KOIVISTO and MARIT SANDSTAD}

\address{Institute of Theoretical Astrophysics, University of Oslo,\\
P.O. Box 1029 Blindern, N-0315 Oslo, Norway\\
$^*$E-mail: yashar.akrami@astro.uio.no}

\begin{abstract}
The recent discovery of a ghost-free, non-linear extension of the Fierz--Pauli theory of massive gravity, and its bigravity formulation, introduced new possibilities of interpreting cosmological observations, in particular, the apparent late-time accelerated expansion of the Universe. Here we discuss such possibilities by studying the background cosmology of the model and comparing its predictions to different cosmological measurements. We place constraints on the model parameters through an extensive statistical analysis of the model, and compare its viability to that of the standard model of cosmology. We demonstrate that the model can yield perfect fits to the data and is capable of explaining the cosmic acceleration in the absence of an explicit cosmological constant or dark energy, but there are a few caveats that must be taken into account when interpreting the results.
\end{abstract}

\keywords{Modified Gravity; Massive Gravity; Bigravity; Dark Energy; Background
Cosmology; Statistical Analysis.}

\bodymatter

\paragraph{}
Recent work~\cite{deRham:2010ik,deRham_Gabadadze_et_Tolley2010,Hassan_et_Rosen2011a,Hassan:2011hr,Hassan_et_Rosen2012a,Hassan_et_al2012a,Hassan_et_Rosen2012b} has solved a long-standing and highly-challenging puzzle in theoretical physics, that was first proposed by Fierz and Pauli in 1939~\cite{Fierz_et_Pauli1939}: ``does a consistent non-linear theory of massive gravity exist?" In Refs.~\refcite{deRham:2010ik} and~\refcite{deRham_Gabadadze_et_Tolley2010}, de Rham, Gabadadze and Tolly showed for the first time that it is possible to give gravity a mass while the theory remains free of the so-called Boulware and Deser ghosts~\cite{Boulware_et_Deser1939} (for a review of massive gravity, see Ref.~\refcite{Hinterbichler2011}). Later, in Refs.~\refcite{Hassan_et_Rosen2011a},~\refcite{Hassan:2011hr},~\refcite{Hassan_et_Rosen2012a},~\refcite{Hassan_et_al2012a} and~\refcite{Hassan_et_Rosen2012b}, Hassan and Rosen demonstrated that such a ghost-free theory can be formulated as a bigravity modification of General Relativity in which the gravity sector is extended by a new metric-like 2-tensor that interacts with the original metric only in particular ways. They also showed that as long as the structure of the theory is preserved, the new field can either be kept non-dynamical or be promoted to a dynamical one.

Not only did the discovery of the ghost-free massive gravity, and its formulation in terms of a bimetric theory, attract the attention of many theoreticians, it also introduced a consistent, very interesting and theoretically well-motivated modification of gravity to the cosmological community (for a review of cosmological applications of modified gravity theories, see Ref.~\refcite{Clifton:2011jh}). Such a non-linear theory of massive gravity may in particular explain cosmic acceleration without introducing the problematic cosmological constant (CC)~\cite{Martin:2012bt} or Dark Energy~\cite{Copeland:2006wr}.

In a longer paper~\cite{Akrami:2012vf}, we investigated the ability of the bigravity theory of massive gravity to address the observed acceleration of the Universe~\cite{Riess:1998cb,Perlmutter:1998np} as an alternative to the $\Lambda$CDM standard model of cosmology. We summarize those findings here; the reader is referred to the longer paper for details of all calculations, numerical and statistical methods used and detailed discussion of the results.

We compare the background predictions of a cosmological model based on the Hassan--Rosen bigravity theory to different cosmological observations. The two metrics of the model are assumed to be spatially flat, homogeneous and isotropic. Only one metric is coupled to matter and is considered as the physical metric that appears in the definitions of all cosmologically interesting observables. The full version of the model possesses six free parameters: $\Omega^0_m$, the present value of the (dark plus baryonic) matter density parameter, and $B_i~(i=0,...,4)$, representing possible interactions between the two metrics (see Refs.~\refcite{Akrami:2012vf} and~\refcite{Strauss_et_al2011} for details). $B_0$ corresponds to the explicit CC for the physical metric. We are therefore mostly interested in sub-models where $B_0$ is set to zero (in order to test the ability of the model to provide a self-acceleration mechanism). We perform an extensive Bayesian parameter estimation procedure using MultiNest algorithm~\cite{Feroz:2008xx} and compute uncertainties around the best-fit points using marginalized posterior probabilities. We in addition employ a semi-frequentist technique to verify whether the parameters are well-constrained or there are correlations between them. We also compare the model's ability to fit the data to that of $\Lambda$CDM by evaluating $\chi^2_{min}$, $p$-value and log-evidence for each model. The data we use include the position of the first peak on the cosmic microwave background angular power spectrum~\cite{Komatsu:2010fb}, the ratio of the sound horizon at the drag epoch to the dilation scale at six different redshifts~\cite{Beutler:2011hx,Percival:2009xn,Blake:2011en}, the luminosity distances to 580 Type Ia Supernovae~\cite{Suzuki:2011hu}, and the present value of the Hubble parameter.

\begin{table}[t]
    \tbl{Best-fit $\chi^2$, $p$-value and log-evidence for the ghost-free bigravity model and its sub-models when constrained by various cosmological measurements at the background level. For each model, parameters that are allowed to vary are marked as ``free''; the non-varying parameters are fixed to zero.}
      {\begin {tabular}{@{}cccccccccc@{}}
        \toprule
        {\bf Model} & ${\bf B_0}$& ${\bf B_1}$ & ${\bf B_2}$ & ${\bf B_3}$ & ${\bf B_4}$ & ${\bf \Omega_m}$& ${\bf \chi^2_{min}}$& {\bf p-value}&{\bf log-evidence}\\\colrule
	${\bf \Lambda}${\bf CDM}& free& 0 & 0 & 0 & 0 & free& \hphantom{0}546.54 & \hphantom{0}0.8709 &-278.50\\
	${\bf (B_1,\Omega_m^0)}$&0& free & 0 & 0 & 0 & free & \hphantom{0}551.60 & \hphantom{0}0.8355 &-281.73\\
	${\bf (B_2,\Omega_m^0)}$&0& 0 & free & 0 & 0 & free & \hphantom{0}894.00 & $<0.0001$ & -450.25\\
	${\bf (B_3,\Omega_m^0)}$&0& 0 & 0 & free & 0 & free & 1700.50 & $<0.0001$ & -850.26\\
        ${\bf (B_1,B_2,\Omega_m^0)}$&0& free & free & 0 & 0 & free & \hphantom{0}546.52 &\hphantom{0}0.8646 & -279.77\\
        ${\bf (B_1,B_3,\Omega_m^0)}$&0& free & 0 & free & 0 & free & \hphantom{0}542.82 & \hphantom{0}0.8878 & -280.10\\
        ${\bf (B_2,B_3,\Omega_m^0)}$&0& 0 & free & free & 0 & free & \hphantom{0}548.04 & \hphantom{0}0.8543 & -280.91\\
        ${\bf (B_1,B_4,\Omega_m^0)}$&0& free & 0 & 0 & free & free & \hphantom{0}548.86 & \hphantom{0}0.8485 & -281.42\\
        ${\bf (B_2,B_4,\Omega_m^0)}$&0& 0 & free & 0 & free & free & \hphantom{0}806.82 & $<0.0001$ & -420.87\\
        ${\bf (B_3,B_4,\Omega_m^0)}$&0& 0 & 0 & free & free & free & \hphantom{0}685.30 & \hphantom{0}0.0023 & -351.14\\
        ${\bf (B_1,B_2,B_3,\Omega_m^0)}$&0& free & free & free & 0 & free & \hphantom{0}546.50 & \hphantom{0}0.8582 & -279.61\\
        ${\bf (B_1,B_2,B_4,\Omega_m^0)}$&0& free & free & 0 & free & free & \hphantom{0}546.52 & \hphantom{0}0.8581 & -279.56\\
        ${\bf (B_1,B_3,B_4,\Omega_m^0)}$&0& free & 0 & free & free & free & \hphantom{0}546.78 & \hphantom{0}0.8563 & -280.00\\
        ${\bf (B_2,B_3,B_4,\Omega_m^0)}$&0& 0 & free & free & free & free & \hphantom{0}549.68 & \hphantom{0}0.8353 & -282.89\\
        ${\bf (B_1,B_2,B_3,B_4,\Omega_m^0)}$&0& free & free & free & free & free & \hphantom{0}546.50 & \hphantom{0}0.8515 & -279.60\\
        {\bf full bigravity model}&free& free & free & free & free & free & \hphantom{0}546.50 & \hphantom{0}0.8445 &-279.82\\\botrule
      \end{tabular}}
      \label{tbl:ResultsDifferentParmaeterRegimes}
  \end{table}

Our numerical results, confirmed with detailed analytical studies, show (see Table~\ref{tbl:ResultsDifferentParmaeterRegimes}) that the model in general gives very good fits to the data at a very high confidence level. Except for $(B_2,\Omega^0_m)$, $(B_3,\Omega^0_m)$, $(B_2,B_4,\Omega^0_m)$ and $(B_3,B_4,\Omega^0_m)$ that are ruled out, other sub-models with $B_0=0$ are in perfect agreement with observations and are statistically as consistent with the data as $\Lambda$CDM (log-evidence should not be considered here because of the prior choices of parameter ranges and prior preferences we may have for a model over the other). In addition, we observe that models with more than one $B$ are degenerate, i.e. parameters are correlated. Such degeneracies can be broken by using more data in a perturbative analysis. Only for $(B_1,\Omega^0_m)$, where we have reliable constraints, the graviton mass can be determined; it is of the order of the present value of the Hubble parameter.

\section*{Acknowledgements}
We thank Tessa Baker, Robert Crittenden, Jonas Enander, Hans Kristian K. Eriksen, Farhan Feroz, Pedro G. Ferreira, Juan Garcia-Bellido, S. F. Hassan, Antony Lewis, Edvard M\"{o}rtsell, David F. Mota, Sigurd K. N{\ae}ss, Claudia de Rham, Rachel A. Rosen, Mikael von Strauss and Andrew J. Tolley for enlightening and helpful discussions. YA is supported by the European Research Council (ERC) Starting Grant StG2010-257080.

\bibliographystyle{ws-procs975x65}
\bibliography{sources}

\end{document}